\documentclass[letterpaper]{article} % DO NOT CHANGE THIS

\usepackage{aaai24}  % DO NOT CHANGE THIS
\usepackage{times}  % DO NOT CHANGE THIS
\usepackage{helvet}  % DO NOT CHANGE THIS
\usepackage{courier}  % DO NOT CHANGE THIS
\usepackage[hyphens]{url}  % DO NOT CHANGE THIS
\usepackage{graphicx} % DO NOT CHANGE THIS
\usepackage{natbib}  % DO NOT CHANGE THIS AND DO NOT ADD ANY OPTIONS TO IT
\usepackage{caption} % DO NOT CHANGE THIS AND DO NOT ADD ANY OPTIONS TO IT
\usepackage{algorithm}
\usepackage{algorithmic}

\usepackage[T1]{fontenc}

\usepackage{newfloat}
\usepackage{listings}
\DeclareCaptionStyle{ruled}{labelfont=normalfont,labelsep=colon,strut=off} % DO NOT CHANGE THIS
\floatstyle{ruled}
\newfloat{listing}{tb}{lst}{}
\floatname{listing}{Listing}
\usepackage{tabularray}
\UseTblrLibrary{booktabs}
\usepackage{subcaption}
\usepackage{multirow}
\usepackage{booktabs}
\usepackage{adjustbox} % adjusting table
\usepackage{bm}
\usepackage[dvipsnames]{xcolor}
\title{
\begin{normalsize}
\begin{center}
    \vspace*{-2cm}
\fcolorbox{red}{white}{Please cite the ICWSM'24 version of this article}
    \vspace{1.2cm}
\end{center}
\end{normalsize}
HyperGraphDis: Leveraging Hypergraphs for Contextual and Social-Based Disinformation Detection}

\author {
	Nikos Salamanos\textsuperscript{\rm 1},
        Pantelitsa Leonidou\textsuperscript{\rm 1},
	Nikolaos Laoutaris\textsuperscript{\rm 2},
	Michael Sirivianos\textsuperscript{\rm 1},\\
        Maria Aspri\textsuperscript{\rm 3},
	Marius Paraschiv\textsuperscript{\rm 2} \\
}
\affiliations {
	\textsuperscript{\rm 1} Cyprus University of Technology, Limassol, Cyprus \\ 
	\textsuperscript{\rm 2} IMDEA Networks, Madrid, Spain \\
        \textsuperscript{\rm 3} LSTECH ESPANA SL, Barchelona, Spain \\
	 nik.salaman@cut.ac.cy, pl.leonidou@edu.cut.ac.cy, nikolaos.laoutaris@imdea.org, michael.sirivianos@cut.ac.cy, \\
  aspri@lstech.io, marius.paraschiv@imdea.org \\

}

\usepackage{booktabs} % for professional tables
\usepackage{makecell}
\usepackage{amsmath}

\usepackage{amssymb}
\usepackage{arydshln}
\usepackage{breakurl}
\usepackage{amsthm}
\usepackage{soul} % for strike-through text
\theoremstyle{definition}

\usepackage[dvipsnames]{xcolor}
\begin{document}

\renewcommand{\arraystretch}{1.1}

\maketitle

\begin{abstract}
In light of the growing impact of disinformation on social, economic, and political landscapes, accurate and efficient identification methods are increasingly critical. This paper introduces HyperGraphDis, a novel approach for detecting disinformation on Twitter that employs a hypergraph-based representation to capture (i) the intricate social structures arising from retweet cascades, (ii) relational features among users, and (iii) semantic and topical nuances. Evaluated on four Twitter datasets -- focusing on the 2016 U.S. Presidential election and the COVID-19 pandemic -- HyperGraphDis outperforms existing methods in both accuracy and computational efficiency, underscoring its effectiveness and scalability for tackling the challenges posed by disinformation dissemination. HyperGraphDis displays exceptional performance on a COVID-19-related dataset, achieving an impressive F1 score (weighted) of approximately 89.5\%. This result represents a notable improvement of around 4\% compared to the other state-of-the-art methods. Additionally, significant enhancements in computation time are observed for both model training and inference. In terms of model training, completion times are accelerated by a factor ranging from 2.3 to 7.6 compared to the second-best method across the four datasets. Similarly, during inference, computation times are 1.3 to 6.8 times faster than the state-of-the-art.
\end{abstract}

\section{Introduction}
\label{sec:introduction}
Social media platforms have evolved from mere fora for informal conversation and idea exchange into essential tools for a diverse array of entities, including non-governmental organizations, businesses, governments, and media outlets. Their ubiquity and ability to personalize content according to user behavior make them pivotal in the modern social landscape. Platforms like Twitter (currently known as X) and Facebook now serve as the principal sources of information on a worldwide scale ~\cite{social_media1}. Nevertheless, the dissemination of false information on these platforms can have detrimental effects on society, politics, and the economy, making disinformation a rapidly escalating cyber threat \cite{cyberthreat}. Contemporary techniques for identifying disinformation on social media chiefly focus on partial data, either concerning user behavior or the topology of the local network. Such methods may lead to unsatisfactory results, given that the effectiveness of disinformation spread relies on a multifaceted interplay of intrinsic elements like user traits and extrinsic factors such as network features and post content.

In this study, we aim to improve disinformation detection accuracy and computational efficiency on social media by introducing an integrated approach that considers both intrinsic and extrinsic variables. For this purpose, we propose HyperGraphDis, an innovative approach utilizing a hypergraph data structure to encapsulate both the content of Twitter retweet cascades and the relational information among users. On Twitter, users create messages (\textit{tweets}), and others share these messages (\textit{retweets}). A posted message initiates a \textit{root-tweet} (the original tweet that gets retweeted), followed by a sequence of retweets forming a tree-like graph known as a \textit{retweet cascade}. The choice of a hypergraph stems from its ability to represent complex relationships more naturally than traditional graphs. Unlike standard graphs with vertices connected by edges, a hypergraph features hyperedges that can connect more than two nodes. This allows for a nuanced representation of Twitter interactions, considering both the content of tweets and retweets (referred to as retweet cascades), as well as the network of users involved in these cascades.

In the initial phase of hypergraph construction, we apply a graph partitioning algorithm to the Twitter social network, where nodes represent users and edges represent social connections. This process effectively identifies user clusters with similar interaction patterns or content sharing. The essence of graph partitioning is to optimize the divisions in such a way that intra-cluster connections are maximized while inter-cluster connections are minimized, thereby identifying natural communities within the Twitter user base.

Upon identifying these user clusters, we undertake a transformation wherein each user in a cluster is substituted with a list of Twitter cascades in which they have partaken. In the hypergraph framework, each previously discovered cluster becomes a hyperedge. The Twitter cascades, which are sequences of tweets and retweets initiated by a root-tweet, are represented as the nodes within these hyperedges. This transformation inherently reshapes the problem space: it turns the complex task of disinformation classification on Twitter from an intricate, multi-variable problem into a more straightforward node classification problem within the hypergraph.

The use of hypergraph data structures offers several advantages over traditional graphs. Unlike conventional graph edges that connect only pairs of nodes, a hyperedge links an entire cluster of users, thereby enriching the relational information in the dataset by capturing multi-way associations. This multi-faceted relational insight becomes even more potent when considering hyperedge intersections, which naturally coincide with the overlap of user clusters. Such overlaps allow for a nuanced understanding of classes of users sharing specific common features, further augmenting the dataset's expressiveness. Moreover, hypergraphs inherently result in less dense network representations, yielding benefits in terms of reduced computational processing times and more efficient storage requirements. This makes hypergraphs an excellent choice for the effective modeling of complex relational systems.

Moreover, we enrich each Twitter cascade node with a set of features, such as the frequency of specific keywords, sentiment analysis scores, or the presence of verified accounts (with the Twitter verification badge). By doing so, we transform the issue of classifying Twitter cascades in terms of disinformation into a node classification task within the hypergraph. For the node classification task, we employ hypergraph neural networks, a modern machine learning approach optimized for hypergraph structures. 

We evaluate our approach on four datasets: (i) an extensive dataset on the 2016 U.S. Presidential Election; (ii) a substantial collection of tweets related to the COVID-19 pandemic, along with social engagement data \cite{mmcovid}; and (iii) the Health Release and Health Story datasets, integral parts of the FakeHealth repository \cite{FakeHealth}. Overall, the HyperGraphDis shows exceptional performance (see Table~\ref{tab:results_metis}). More specifically, evaluating it with MM-COVID achieves an impressive F1 score of around 89.5\%. It outperforms the Meta-graph method \cite{ours} by approximately 4\%, and Cluster-GCN \cite{cluster-gcn} by 33\%. Additionally, HyperGraphDis outperforms HGFND \cite{Ujun:2022} by 6.4\% and other state-of-the-art ML text-based techniques by 10\% (see Table~\ref{tab:other_solutions}).
Furthermore, noteworthy enhancements are observed in the computation time for both model training and inference processes. Regarding model training, the completion time is notably expedited, ranging from 2.3 times to 7.6 times faster than the second-best method per dataset (in terms of F1 score). Similarly, computational time exhibits increased efficiency during inference, ranging from 1.3 times to 6.8 times faster than the second-best method across the four datasets.

The results demonstrate that our method not only yields levels of accuracy that are comparable to or exceed existing methods relying on both tweet content and user relationships, but it also does so while offering the advantages of reduced data size and faster classification run times. Moreover, the inherent design of our method makes it highly scalable, allowing it to adapt to varying sizes and complexities of data. 

\noindent\textbf{Data availability} Part of the US Election dataset is publicly available~\cite{US2016Election} under proper restrictions for compliance with Twitter's Terms of Service (Tos) and General Data Protection Regulation (GDPR). The FakeHealth repository and the MM-COVID dataset are publicly available~\cite{FakeHealth}, \cite{mmcovid}.

\section{Related Work}
\label{sec:related_work}
Considerable research has been dedicated to the problem of identifying false information on social media platforms. Initial methods primarily focused on feature-based detection at the individual user level. The work by \cite{ngram}, for instance, employs a variety of feature extraction techniques on social media content to identify linguistic patterns indicative of disinformation. However, this approach fails to account for the relational data between the information sources and the users on platforms like Twitter. Similarly, \cite{text_style} decompose the problem into two distinct phases: feature extraction and classification, while \cite{lecunn} employ character-level convolutional neural networks to merge these steps. More contemporary methods include the use of BERT language models for feature extraction as outlined by \cite{fakebert} and reviewed in depth by \cite{fakenews_review}. On another front, \cite{castillo} evaluate the trustworthiness of social media posts by examining users' historical behavior, the tweets' content, and additional external data. \cite{shu} take a different tack by extracting features from individual user profiles to identify those more likely to disseminate false information. This bears resemblance to the strategies discussed in \cite{user_based1, user_based2, user_based3}, which leverage characteristics of user types to categorize posts as false or genuine.

Turning to methods that incorporate social context, these consider both the relationships between users sharing news and additional contextual information about the user and the news content. \cite{event_credibility} introduce an approach that evaluates credibility through a network that includes events, tweets, and users. While this conceptually aligns with our method, it overlooks the complex structural data of information cascades and their similarities. \cite{subnet} suggest a co-attention mechanism between sentences and comments to explain why a news item is flagged as false by an algorithm. Concurrently, \cite{tracing} focus on identifying false information by analyzing its propagation behavior across the network.

A method closely related to ours is discussed by \cite{similar_to_us}, where an embedding framework known as TriFN is utilized. It models the relationships between publishers and news items, as well as user interactions with those news pieces. However, it lacks the incorporation of relational data between events or cascades, which can be crucial for evaluating historical user behavior. Additionally, it does not leverage semantic features gained through topic modeling and sentiment analysis. Another noteworthy method is presented by \cite{hierarchical}, who construct a hierarchical propagation network for false information and engage in a comparative study between false and true news from linguistic, structural, and temporal lenses.

We mention here other recent efforts, such as \cite{new1}, which employs automatic methods and machine learning models like RoBERTa-large to assess the veracity of claims in medical datasets. This work complements efforts in understanding web-based relationships, where \cite{new2} utilizes Common Crawl data to analyze hyperlinks between websites. In \cite{new3}, the authors present a comprehensive framework to study Reddit communities, targeting misinformation, toxicity, and political polarization. Similarly, \cite{new4} delve into the behavior of social bots on Twitter, categorizing them based on their political leanings during the 2018 U.S. midterms.\cite{new5} extend this by introducing metrics, MRD and MRS, to quantify user migrations across subreddits on the Reddit social platform. Furthermore, 
In \cite{new6}, the focus is on detecting repurposed Twitter accounts using supervised learning, shedding light on their characteristics and behaviors.

In \cite{sota-fake-news-detection}, the authors conducted a meta-analysis of state-of-the-art Machine Learning techniques for text-based fake news detection, including Random Forest, CNN-based, LSTM-based, and BERT-based models. They evaluated these models on benchmark datasets, including MM-COVID. We evaluated the HyperGraphDis method using MM-COVID, consistently achieving superior performance compared to all mentioned models in terms of F1 score. 
The SOMPS-Net framework~\cite{somps-net}, with the Social Interaction Graph and Publisher and News Statistics components, achieved a 72.7\% accuracy and 79.6\% F1 score on the Health Story dataset for fake news prediction. Evaluating HyperGraphDis on the same dataset, we observed an accuracy of 75.2\% and F1 score of 72.2\%. We attribute this decline in F1 score to the inherent characteristics of the cascades within the Health Story dataset, which tend to be either extremely small or predominantly structured like star-like graphs. This characteristic leads to embedding vectors that exhibit high levels of similarity, consequently affecting the overall effectiveness of our method.

The methods most closely related to ours are \cite{ours} and \cite{Ujun:2022}. In \cite{ours}, the Meta-graph method captures content and relational information. \cite{Ujun:2022} introduces a hypergraph-based neural network (HGFND) for detecting relational fake news with limited labeled data. They formulate fake news detection as a semi-supervised node classification task in a hypergraph, capturing group-level interactions among news articles. Our analysis reveals that HyperGraphDis outperforms these methods, providing superior results with enhanced computational efficiency.

In evaluating the performance metrics of our proposed method, several key factors underscore its superiority, particularly when compared to classical graph-based approaches. First, the inherently reduced connectivity of the hypergraph structure yields faster computational times and minimizes hardware resource requirements. This efficiency is not achieved at the expense of informational richness; to the contrary, hyperedges encapsulate a wealth of multi-way relationships among nodes, providing a more nuanced understanding of underlying data patterns. Additionally, the intersections of hyperedges serve as focal points for identifying user subsets with highly detailed common features, adding an extra layer of depth to the dataset's relational information content.

% %%%%%%%%%%%%%%
\begin{figure*}[t!]
  \centering
  \includegraphics[width=0.95\linewidth]{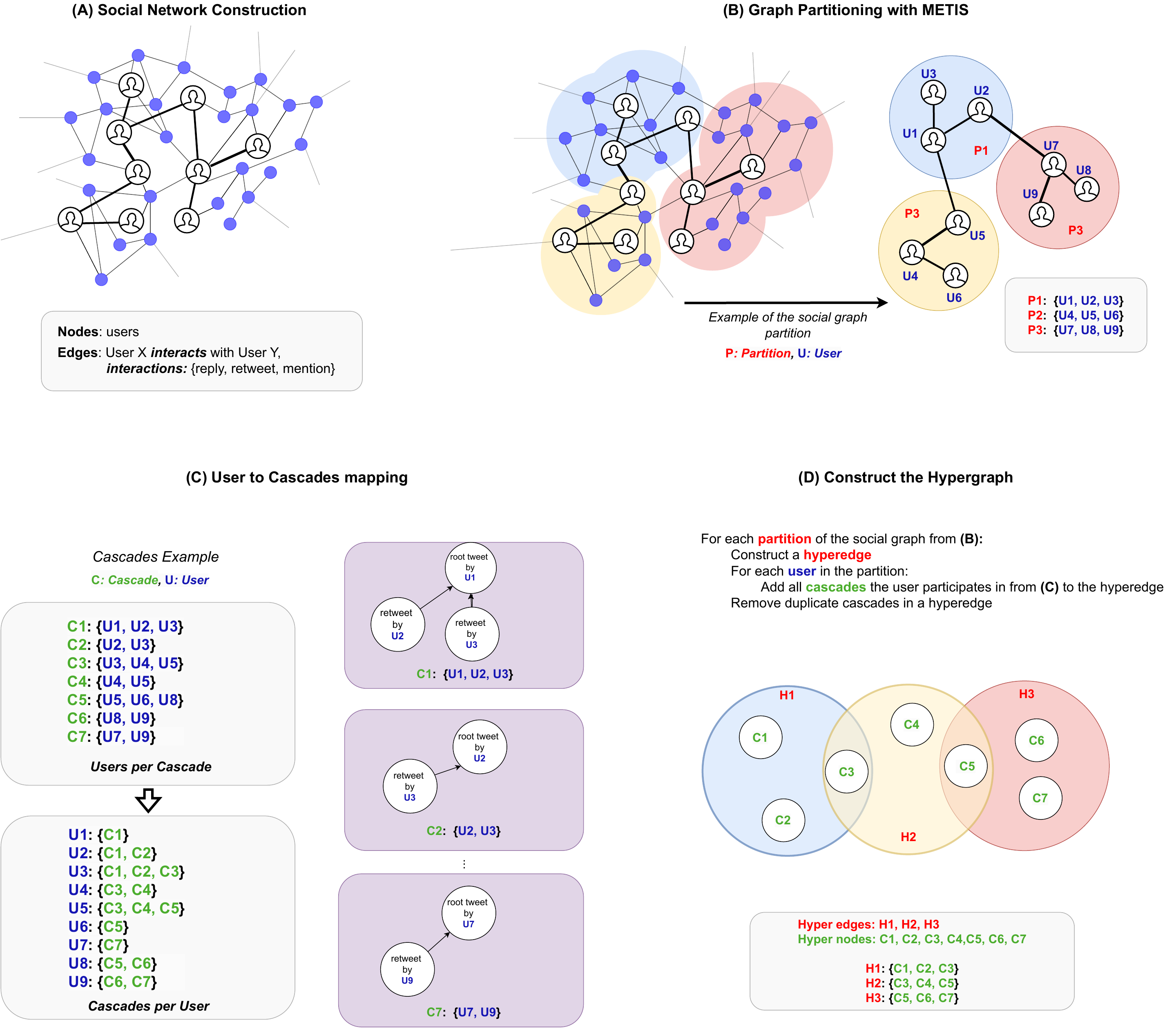}
	\caption{A toy example of the proposed hypergraph construction pipeline.}
	\label{fig:hypergraph_pipeline}
\end{figure*}
% %%%%%%%%%%%%%

\section{Similar Approaches}
\label{sec:approaches}

\subsection{Meta-graph}
In ~\cite{ours}, the authors introduce a \textit{meta-graph} data structure to represent events in a social network. Formally, a meta-graph \(\mathcal{G} = (V, E)\) consists of vertices \(V\) and edges \(E\), where each vertex corresponds to a social network event and is characterized by a feature vector \(X_{v}\).

The node feature vector \(X_{v}\) encapsulates user attributes, tweet content, and cascade structural information. Specifically, \(X_{v} = (C_{\text{emb}} || X_u || T_{\text{emb}} || S \ldots)\), where \(||\) signifies vector concatenation. The term \(C_{\text{emb}}\) denotes the cascade embedding vector, \(X_u\) encompasses concatenated user features, \(T_{\text{emb}}\) is the text embedding vector, and \(S\) represents sentiment analysis scores if available~\cite{DeepWalk, bert}.

Edges in the meta-graph are constructed based on common users or topics and are described by edge feature vectors \(R_{e_{i,j}}\). An example vector is \(R_{e_{i,j}} = (N_{u_{i,j}} || V_{i,j} || H_{i,j} \ldots)\). Here, \(N_{u_{i,j}}\) indicates the number of common users between cascades \(i\) and \(j\), \(V_{i,j}\) is a graph similarity metric, and \(H_{i,j}\) is a content similarity metric between the root-tweets of the two cascades~\cite{graph_sim1, graph_sim2, graph_sim3, graph_sim4, text_sim}. Both node and edge feature vectors are designed for extensibility ease, allowing the incorporation of additional relational information when needed.

Despite its utility in certain scenarios, the meta-graph method presents several notable limitations that hamper its broader applicability. One primary drawback is the high density of its graph representation, which demands considerable computational resources and results in an increased hardware footprint. Additionally, the meta-graph edges are confined to capturing bipartite connections between entities, failing to represent more complex relational information. This restrictiveness also impedes the exploitation of partial feature overlaps among different user clusters. To overcome these challenges, we introduce HyperGraphDis, a novel approach that employs a hypergraph data structure to address these inherent shortcomings effectively.

\subsection{HyperGraph for Fake News Detection (HGFND)}
\label{subsec:HGFND}
In \cite{Ujun:2022}, the authors tackle the challenge of detecting fake news by framing it as a semi-supervised node classification task within the context of a hypergraph. They propose a hypergraph construction methodology that establishes connections between news pieces (nodes) based on both the news content and social context – the propagation of news on Twitter. They create a hypergraph with three hyperedge types: \textbf{i) User:} a hyperedge for each user I.D., linking news pieces shared by a specific user; \textbf{ii) Time:} a hyperedge for a proximal time range, connecting news pieces shared during the same period; \textbf{iii) Entity:} a hyperedge for each entity, connecting news pieces that share the same content entity (e.g., "Covid-19").

Finally, they introduce HGFND, a hypergraph neural network model designed to handle noisy relations within the hypergraph. HGFND employs a dual-level attention mechanism, assessing the significance of news relations at both node and hyperedge levels. Experimental results on two benchmark datasets (Politifact and Gossipcop) demonstrate HGFND's high performance in the news classification task. We evaluate the performance of HGFND in the cascade classification task with the hypergraph constructed following our approach (see Section \textit{The HyperGraphDis Method}) and the results are shown in Table~\ref{tab:results_metis}.

\subsection{Cluster-GCN}
\label{subsec:cluster-gcn}
In~\cite{cluster-gcn}, the authors present Cluster-Graph Convolutional Network (Cluster-GCN) that leverages graph clustering to achieve optimized Stochastic Gradient Descent (SGD)-based training in a mini-batch fashion. Cluster-GCN first partitions the graph utilizing METIS \cite{metis} -- a graph partitioning algorithm, to derive non-overlapping clusters of nodes. As a mini-batch of size \textit{q}, it randomly selects \textit{q} clusters to construct a subgraph with nodes from the chosen clusters and the links connecting them in the original graph. 

For each SGD update on a mini-batch, this approach constrains GNNs to convolve exclusively within their designated subgraphs, thereby addressing the issue of neighborhood expansion. This restriction results in optimized memory consumption and reduced computational costs during training, allowing large-scale training on graph structures. 
In this work, we evaluate the performance of Cluster-GCN on the meta-graph data structure as a baseline. Note that in Cluster-GCN, METIS is applied to the meta-graph to generate clusters of cascades, whereas in HyperGraphDis, METIS generates clusters of users applied to the social network.

\section{The HyperGraphDis Method}
\label{subsec:convolpergraph_construction}

The HyperGraphDis method addresses the detection of disinformation in social networks by exploiting network structure as well as post content and user sentiment. As we will show in the next section, the network structure has a clear impact on classification results. As part of the HyperGraphDis method, a hypergraph is constructed from the user-user social network as well as the individual tweet-retweet cascades, which will serve as the nodes of the hypergraph. The hypergraph structure is used to perform binary classification of retweet cascades into fake and not-fake classes. Two primary challenges need to be tackled to apply our methodology to the raw Twitter data. First, the actual structure of the social network, specifically the follower graph, remains largely unknown, as access to users' follower lists is not always available. Second, the data obtained through the Twitter API offer limited insights into the origins of influence within retweet cascades. The only accessible information consists of the root-tweet and the root-user associated with a particular retweet. In simpler terms, all retweeters are linked to the root user, creating a star-like cascade structure that does not always accurately represent user-to-user interactions or the source of information diffusion. The methodology steps to address the above issues are the following:

\subsection{Phase-1: The Hypergraph Construction}
\label{subsec:hypergraph_construction}

Given a large collection of tweets, we construct a hypergraph with the retweet cascades as nodes (i) From the overall collection of tweets (tweets, replies, retweets), we construct the user-user social network -- the graph of interactions between the users. This graph represents an approximation of the Twitter social network. (ii) We partition this graph (in disjoint sets of nodes/users) by applying METIS -- a state-of-the-art graph partitioning algorithm. Then, in each partition (set of users), we replace the users with the retweet cascades that have participated. In simple words, clusters of cascades replace clusters of users, and each of these clusters represents a hyperedge.

Figure~\ref{fig:hypergraph_pipeline} presents a toy example of the main HyperGraphDis phases for the hypergraph construction where the classification of disinformation is taking place. The hypernodes are the cascades that are represented by feature vectors (as defined in the next Section, ``Phase-2: The Hypernodes Features Vectors''). In our methodology, a hyperedge connects two or more cascades if the users of the same partition of the Twitter social graph participate in these cascades. To illustrate this, consider the social network and the corresponding graph partition in Figure~\ref{fig:hypergraph_pipeline}(A) \& (B). Initially, we create a user-to-cascades mapping, as shown in Figure~\ref{fig:hypergraph_pipeline}(C). Subsequently, we replace each user within a given partition from Figure~\ref{fig:hypergraph_pipeline}(B) with the cascades they have engaged with from Figure~\ref{fig:hypergraph_pipeline}(C). This process leads to the formation of the hypergraph (Figure~\ref{fig:hypergraph_pipeline}(D)) where the hyperedges connect these sets of cascades. The construction of the hypergraph is scalable since several types of cascade connections (hyperedges) can be defined. For example, we can construct hyperedges based on tweet hashtags; i.e., two or more cascades belong to the same hyperedge if the root-tweets have the same hashtag. 

In Figure~\ref{fig:hypergraph_pipeline}, we illustrate the process of building the cascades hypergraph. This begins with the formation of the social network and follows through to capturing users' participation in cascades, all based on the selected Twitter dataset under consideration. A detailed description of the proposed pipeline is as follows:

% %%%%%%%%%%%%%%%%%%%%%%%%
\begin{figure}[t!]
  \centering
  \includegraphics[width=0.95\linewidth]{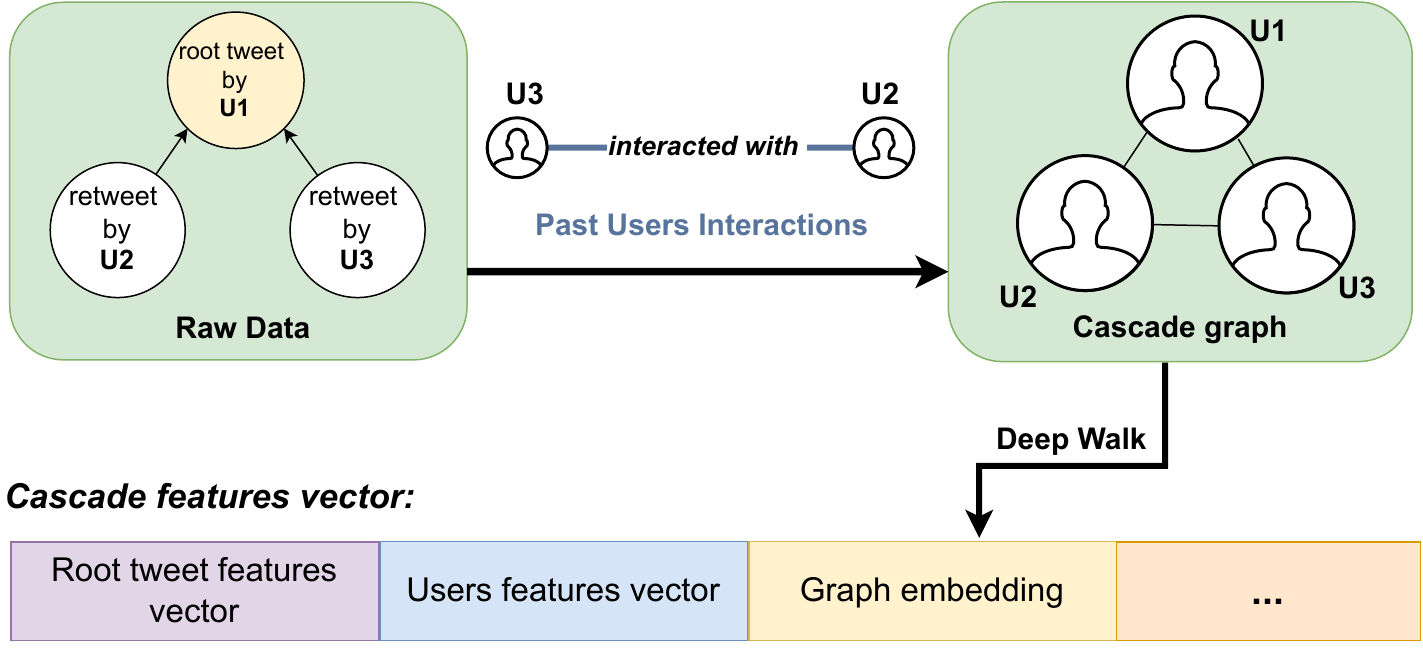}
	\caption{A toy example of a hypergraph node. Left-hand: the raw retweet data provided by Twitter API, always corresponds to a star-like graph with limited structural information. Right-hand: We enhance the raw data by appending the past interactions between the retweeters (i.e., the users) to the star-like graph. Finally, we compute the node embeddings (in this enhanced subgraph) using the DeepWalk algorithm. Bottom: The final feature vector consists of the DeepWalk embeddings together with users' features and tweet-text features (topics, sentiments, etc.)}
	\label{fig:cascades_vector_construction}
\end{figure}
% %%%%%%%%%%%%%%%%%%%%%%%%

\textbf{(A) Social Network Construction} Initially, we create an approximation of the Twitter social graph, also known as the follow-graph, grounded in the activities of users within our dataset. Within this social network, each user corresponds to a node. We add a directed edge for any interaction between two nodes. The ``interactions'' are defined by the user-user actions in the Twitter platform, such as ``retweet,'' ``reply'', or ``mention''. Hence, an edge(\textit{i,j}) between two nodes implies that the \textit{node i} follows the \textit{node j}. The output of this process is a multi-graph because of the possibility of multiple interactions (edges) between two nodes. We remove the duplicate edges and keep the earliest ones. 

\textbf{(B) Graph Partitioning} Having the resulting Twitter social graph, we conduct graph partitioning to identify communities or groups of users. After applying the selected graph partitioning technique, we get a list of users per partition. In this paper, we leverage the \textbf{METIS} graph partitioning algorithm \cite{metis}.

\textbf{(C) User to Cascades Mapping} We define a list of users that are involved in each cascade based on the raw Twitter data. Finally, we derive a user-to-cascade mapping that denotes that a given user $i$ participated in the set of cascades \{x, y, z\}.

\textbf{(D) The Final Hypergraph} Using the graph partition from Step-(B) and the mapping user-to-cascade from Step-(C), we replace the users in the partition clusters with the list of cascades they participated in. This process results in sets of cascades (removing the duplicates), which will define the hyperedges of the final hypergraph.

\subsection{Phase-2: The Hypernodes Features Vectors}
\label{sec:features_construction}
A retweet cascade is a series of chain events upon the same root-tweet where users have retweeted the root--tweet, whereas some others have retweeted a retweet of a friend (retweet of a retweet). This tree-like structure reflects who was influenced by whom during the retweeting process. Unfortunately, the raw data by Twitter API does not provide that since it contains only the root-user and retweeter IDs. In short, the raw data represents a star-like graph where all retweeters are only connected to the root user. This form lacks sufficient structural information.

To tackle this issue, we construct a subgraph comprising the star-like graph along with past interactions involving the users who have participated in the cascade in question. Specifically, for a given retweeter $i$ who engaged in the retweet action at date $t$, we determine their pre-date $t$ connections among their friends and ascertain which of them are among the retweeters in this cascade. Subsequently, we augment this collection of edges, if present, with the star-like configuration where each retweeter maintains a connection with the root user. Finally, we eliminate any duplicate edges and transform the resulting graph into an undirected representation. Through this methodology, the subgraph consistently retains its connectivity.

The rationale of this approach is that each cascade (i.e., a node of the hypergraph) should have the corresponding subgraph structure as a feature that will reflect the social interactions that the retweeters had. For this purpose, we apply the DeepWalk algorithm~\cite{DeepWalk} to produce the embedding of a cascade subgraph to a low dimensional space. For the implementation, we used the \textit{Karate Club} extension for the NetworkX~\cite{karateclub}. The DeepWalk embedding is a $N \times 128$ matrix, where $N$ is the number of retweeters. The default parameters are: (i) $\text{Number of random walks}=10$; (ii) $\text{Length of random walks}=80$ and for embeddings (iii) $\text{Dimensionality of embedding}=128$.

Figure~\ref{fig:cascades_vector_construction} provides a toy example: (i) Left-hand figure: the raw data that are provided by the Twitter API correspond to a star-like graph structure with limited structural information regarding the user-user relation between the retweeters. (ii) Right-hand figure: The augmented cascade graph. (iii) Bottom figure: the final cascade object -- a set of features that represent the retweet cascade. The feature vector consists of the individual users' features, the DeepWalk embeddings as well as topic and sentiment analysis (as the tweet-text features). This will be part of the training data of HyperGraphDis.

\subsection{Phase-3: Cascade Classification}
\label{sec:pipeline_classification}
We formulate the cascade classification problem as a node classification task. The HyperGraphDis uses a Hypergraph Convolution (HypergraphConv) layer~\cite{hypergraphconv} followed by three linear (fully connected) layers. Given a hypergraph structure from Phase 1 and 2, the nodes, their feature vectors, and the hyperedge index serve as input to the HypergraphConv layer, where information aggregation occurs within hyperedges, considering the collective influence of interconnected nodes. The output is a set of transformed node features representing the updated state of each node. The architecture employs Rectified Linear Unit (ReLU) activation functions for non-linearity after each layer. We add dropout regularization between linear layers to enhance generalization. The output of the pipeline represents the model's predictions (i.e., cascade is fake or non-fake). While training Graph Neural Networks (GNNs), it is crucial to highlight the absence of label leakage issues. This is attributed to the fact that, during the training phase, the labels of nodes beyond the training dataset remain unknown. This ensures the integrity of the training process, preventing unintended information leakage from labels outside the designated training set. 

%%%%%%%%%
\begin{table*}[h!]
\begin{center}
\begin{small}
\adjustbox{max width=\textwidth}{%
%\resizebox{\textwidth}{!}{%
\begin{tabular}{lcc|cccc cc|cc}
\toprule
        & \multicolumn{2}{c|}{{Retweet Cascades}} & \multicolumn{6}{c|}{{Social Network}} & \multicolumn{2}{c}{{Meta-graph}} \\
\toprule
        Dataset &\#fake & \#non-fake & \makecell{\#nodes\\(users)} & \#edges & \makecell{\#connected\\components} & \makecell{Giant Component\\ \% of nodes} & \makecell{Infomap\\\#comm} & \makecell{Louvain\\\#comm} & \makecell{\#nodes\\(cascades)} & \#edges \\
\midrule
US Election     & 5,890   & 2,433 &   482,171   & 34,691,548 & 1 & 100\%       &  2047 &  14    & 8,319 & 7,763,018  \\ 
Health Story    & 3,574 & 9,689 &   147,607   &   187,096  & 12,719 & 73.75\%  &   12894     & 12850     &  13,263 & 100,001     \\ 
MM-Covid    & 578 & 1,230     &   38,813    & 50,688     & 4 & 99.78\%         &   47    &   101   & 1,808    &   87,948  \\ 
Health Release  & 929 & 1,040  & 18,470      & 21,502     & 1,595 & 73.32\%    &    1645   &   1659   & 1,969 & 11,581 \\
\bottomrule
\end{tabular}
}
\end{small}
\end{center}
\caption{Cascades, social network and meta-graph statistics per dataset.} 
\label{tab:social-meta-graph-statistics}
\end{table*}

%%%%%%%%%%
\begin{table}[ht]
\center
\begin{small}   
  \begin{tabular}{p{3cm}|p{1cm}p{0.9cm}p{0.9cm}}
  \hline
   \textbf{Features} & \textbf{US Election} & \textbf{MM-COVID} & \textbf{Fake Health} \\
     \hline    
    \multicolumn{4}{l}{\textbf{User-related}} \\
    \hline
    DeepWalk Embedding & \checkmark & \checkmark & \checkmark  \\
    Account Creation Date & \checkmark & \checkmark & \checkmark  \\    
    Verified Account & \checkmark & \checkmark & \checkmark \\
    Account Language & \checkmark &  \checkmark & \checkmark \\
    Max Statuses count & \checkmark &  & \checkmark  \\
    Max Favorites count & \checkmark &  & \checkmark  \\
    Max listed count & & \checkmark &  \\
    Reaction Time & \checkmark & \checkmark &   \\
    Max Followers count & \checkmark & \checkmark & \checkmark  \\
    Max Friends count & \checkmark & \checkmark & \checkmark  \\
    Max impression count & & \checkmark &  \\
    Max reply count & & \checkmark &  \\
    Max quote count & & \checkmark & \\
    Max like count  & & \checkmark & \\
    
    \hline
    \multicolumn{4}{l}{\textbf{Text-related}} \\
    \hline
    Sentiment Analysis &\checkmark & \checkmark &\checkmark \\
    Topic Detection & \checkmark &  & \checkmark  \\
    \hline
  \end{tabular}
\caption{Cascade features.}
\label{tab:cascade-features}
\end{small}
\end{table}
%%%%%%%%%%%%%%%%%%%%%%%%%%%%%%%%%%%%%%%%%%%%%%

\section{Dataset Descriptions}
\label{sec:social_media_data}

\noindent\textbf{US Election}
The dataset in \cite{US2016Election}, encompassing 152.5 million tweets from 9.9 million users during the 2016 U.S. presidential election, was harvested between September 21st and November 7th, 2016, using the Tweepy Python library. Each tweet is feature-rich, encapsulating 27 attributes related to the tweet and the corresponding Twitter account, including tweet-ID, tweet-text, user-ID, screen name, and more. Additional metadata, such as mentions and embedded URLs, are also part of the dataset. The dataset also emphasizes retweet cascades, specifically those rich in transmitted information and broad user engagement. This includes cascades where the original tweet contains at least one web or media URL and has been retweeted by a minimum of 100 users. In total, 46.4K such retweet cascades, involving 19.6 million tweets, are included for analysis. With labeled media and web URLs -- embedded in the tweets -- categorized as "fake," "non-fake," or "unknown," derived through a multi-step methodology and manual annotation using Fact-Checking tools, the dataset provides a well-curated set of 43.9K URLs. There are 6525 URLs labeled as "fake" and "non-fake" that are present in 8323 cascades (see Table~\ref{tab:social-meta-graph-statistics}). 

\noindent\textbf{MM-COVID}
The authors in \cite{mmcovid} created MM-COVID, a multilingual and multimodal dataset to combat misinformation on COVID-19. Despite the news content, the dataset also contains Twitter's social network and spatial-temporal information. The dataset consists of approximately 4K fake news articles and about 7K trustworthy articles related to COVID-19 across six languages. There are over 1K fake news documents in the English language subset and approximately 5K legitimate articles. We dehydrated the user engagement data only for the English language dataset subset using Twarc\footnote{\url{https://twarc-project.readthedocs.io/en/latest/}} (a Python library for collecting data via Twitter API). Only a subset of the data was available to collect at the time of the dehydration of the existing dataset.  
In summary, we used 53,869 tweets, 10,387 replies, and 85,121 retweets by 93,152 users for the experimental evaluation. From this dataset, we kept the 1,808 cascades that have at least 10 participants (see Table~\ref{tab:social-meta-graph-statistics}). The tweets are in English only; the quoted tweets have been discarded.

\noindent\textbf{FakeHealth}
To rigorously assess the efficacy of our proposed methodology, we employ two datasets originating from the FakeHealth repository~\cite{FakeHealth}. While the authors of the repository are constrained by Twitter's user privacy policies from publicly disseminating the full extent of user engagement and network data, they have circumvented this limitation by offering an API~\url{https://github.com/EnyanDai/FakeHealth}. This API facilitates the acquisition of comprehensive data pertaining to users' social engagements and networks. Using this API, we collected the following datasets: \textbf{(i) Health Release} with a size of 60,006 -- 43,245 tweets, 1,418 replies, and 15,343 retweets; \textbf{(ii) Health Story} with a size of 487,195 -- 357,851 tweets, 23,632 replies, and 105,712 retweets. It is noteworthy that the final data metrics we report exhibit a marginal reduction, approximately 1.1\%, compared to the figures disseminated by the original authors. Several factors contribute to this discrepancy: (i) modifications in user settings that restrict public accessibility to their tweets, (ii) deletions of tweets by Twitter in adherence to their internal policies, and (iii) the outright deletion of Twitter user accounts.

\section{Evaluation of Performance Benefits}
\label{sec:experiments}
In this section, we evaluate HyperGraphDis across all four previously mentioned datasets and simultaneously compare it with the Meta-graph, HGFND and Cluster-GCN approaches. First, we present the basic structures across the four datasets that are the input of the HyperGraphDis and the other three methods, and then we discuss the actual classification performance.

The code has been implemented in Python and executed in the CPU of an ASUS laptop with Intel Core i7-6700HQ CPU \@ 2.60GHz × 8; 32GB RAM; Ubuntu 22.04.3 LTS. The HypergraphConv~\cite{hypergraphconv}, GCNConv~\cite{gcn} and Cluster-GCN~\cite{cluster-gcn} have been implemented in PyTorch Geometric \cite{pyg}, maintaining the following hyperparameters constant across experiments: $\text{\#epochs} = 100$, $\text{dropout}=0.5$, $\text{learning rate} = 5*10^{-4}$. For the HGFND we use the official implementation\footnote{\url{https://github.com/ujeong1/IEEEBigdata22_HGFND}}. We use the default parameters $\text{dropout}=0.3$ and $\text{learning rate} = 0.001$. In all experiments, we train the models for 100 epochs. We use an 80/20 train-test split on the labeled cascade set across all datasets and methods. We repeat the experiments for five trials, where we regenerate the train-test split randomly in each trial. Finally, we use the official implementation of Infomap\footnote{\url{https://www.mapequation.org/infomap/}} executing multi-level clustering for 10 internal trials. Louvain has been implemented by NetworkX\footnote{\url{https://networkx.org}}.

\noindent\textbf{Cascades, Social Networks and Meta-graphs} The statistical characteristics of the cascades, social networks, and meta-graphs constructed for the four datasets are presented in Table~\ref{tab:social-meta-graph-statistics}. In the US Election dataset, since the meta-graph constructed is extremely dense (15,946,910 undirected edges connect 8323 nodes/cascades), we applied the disparity filter~~\cite{disparity_filter} for $\alpha = 0.5$, which depicts the best classification results according to~\cite{ours}. 49\% of the edges survived the significance test. However, 4 nodes remained isolated and were discarded, resulting in a meta-graph of 8319 cascades.

\subsection{Cascade Features}
In all datasets, we define a threshold for the number of initial retweeters (users) that we will include in the features (to prevent out-of-memory issues). The thresholds are US Election: 250; MM-COVID: 50; FakeHealth: 60. After constructing the feature matrix, we apply dimensionality reduction using PCA. The dimensions are reduced to US Election: 90, MM-COVID: 60; FakeHealth: 60. Table~\ref{tab:cascade-features} presents the selected parameter values and outlines the user and text-related features employed in the creation of cascade features for each dataset. Note that certain features are not available for certain datasets. 

\noindent\textbf{User-related Features} The user-related features of a given node/cascade with $N$ users (the retweeters plus the root--user) may include the following: The DeepWalk embedding of the correspondent subgraph -- a $N \times 128$ matrix; The dates of users' account creation; The maximum value of users' followers count; The maximum value of users' friends count; The maximum value of users' statuses count; The maximum value of users' favorites count; The maximum value per user and metric across all the tweets that she has posted: listed count; impression count; reply count; quote count; like count; A Boolean identifier whether the users' account is verified; The language of the users' account; users' ``reaction time'', namely, the time difference between a retweet and the root-tweet. 

\noindent\textbf{Text-related Features} In all datasets, we have included the sentiment \textit{(label, score)} of the root--tweet -- two values, either $(2, score)$ for ``Positive'' label or $(1, score)$ for ``Negative'' label. We use the Transformers\footnote{\url{https://huggingface.co/docs/transformers}} library for the sentiment analysis. In addition, we performed topic detection for the \textbf{US Election} dataset only, following the approach of~\cite{ours}. For this purpose, we employ a Hierarchical Dirichlet Process (HDP) \cite{hdp1, hdp2}, using the Tomotopy\footnote{\url{https://bab2min.github.io/tomotopy/v0.12.6/en/}} library. Then, we include the most representative topic of the root--tweet in the cascades' feature vector based on the three topic models (the three versions of the HDP model). For the two \textbf{FakeHealth} datasets, we set as topic (per cascade) the news' ``tags''. This information is provided by the ``reviews'' in the FakeHealth. 
 
%%%%%%%%%%%%%%%%%%%%%%%%%%%%%%%%%%%%%%%%%%%%%%%%%%%%%%%%%%%%%%%%%%
\begin{table*}[ht!]
\small
\begin{center}
\begin{small}
\adjustbox{max width=\textwidth}{%
\begin{tabular}{|lcl | cccc | cccc|}
\hline
\multicolumn{3}{|l|}{\textbf{Methods}} & \multirow{2}{*}{Accuracy (\%)} & F1 Score (\%)  & Train & Inference &  \multirow{2}{*}{Accuracy (\%)}    & F1 Score (\%)  & Train & Inference  \\

  &    \multicolumn{2}{c|}{Graph Partitioning}     &     &  (weighted) &  Time (sec) & Time (sec) &     &  (weighted) &Time (sec)& Time (sec) \\  \cline{4-11} 

& \multicolumn{2}{c|}{(method: \#clusters)} & \multicolumn{4}{c|}{\textbf{2016 US Election}}& \multicolumn{4}{c|}{\textbf{Health Story}} \\
 \cline{1-11} 

\multirow{ 11}{*}{HyperGraphDis} & METIS: & 20 &  $82.86{\pm}0.55$ &  $82.76{\pm}0.54$ &           $224.45$ &            "    $0.681$ &  $73.04{\pm}0.54$ &   $64.62{\pm}1.5$ & $71.41$ &                $0.249$ \\
& " &50   &  $84.66{\pm}0.44$ &  $84.47{\pm}0.44$ &           $447.27$ &                 $1.59$ &  $73.74{\pm}0.64$      &  $67.83{\pm}0.95$ &            $72.81$ &                $0.214$ \\
& " &100  &  $84.82{\pm}0.93$ &  $84.62{\pm}0.96$ &           $703.98$ &                $2.254$ &   $74.36{\pm}0.3$      &  $67.98{\pm}0.56$ &            $74.14$ &                $0.246$ \\
& " &500  &  $85.96{\pm}0.34$ &  \bm{$85.73{\pm}0.35$} & $1704.78$ & $5.212$     &  $74.97{\pm}0.52$      &  $70.58{\pm}0.82$ &            $91.46$ &                $0.269$ \\
& " &1000 &   $85.6{\pm}0.69$ &  $85.26{\pm}0.78$ &           $2037.5$ &                 $6.88$ &  $75.21{\pm}0.83$ &  \bm{$72.23{\pm}0.92$} &    $94.14$ &                $0.268$ \\
&" &2000 &  $85.35{\pm}0.83$ &   $85.01{\pm}0.9$ &          $2472.29$ &                $7.972$ &  $73.77{\pm}0.78$      &  $70.92{\pm}0.85$ &           $115.09$ &                $0.361$ \\
&" &3000 &   $85.9{\pm}0.75$ &  $85.51{\pm}0.76$ &          $2740.06$ &                $9.202$ &  $73.23{\pm}0.99$      &   $70.94{\pm}0.7$ &           $119.81$ &                $0.356$ \\
&" &4000 &   $85.8{\pm}0.87$ &  $85.37{\pm}0.92$ &          $2914.75$ &                $9.856$ &  $72.91{\pm}0.69$      &  $70.95{\pm}0.84$ &           $126.15$ &                 $0.37$ \\
&" &5000 &  $85.57{\pm}0.44$ &  $85.23{\pm}0.47$ &          $3036.64$ &                $10.29$ &  $73.05{\pm}0.95$      &  $70.36{\pm}0.69$ &           $125.17$ &                $0.434$ \\

& Infomap: &* &  $78.37{\pm}0.88$ &  $77.17{\pm}0.81$ &  $113.77$ &  $0.379$ &        $75.56{\pm}0.49$ &  $70.41{\pm}0.7$ &            $57.36$ &                $0.212$ \\

& Louvain: & * &  $77.53{\pm}0.56$ &  $74.77{\pm}0.87$ &   $41.01$ &  $0.184$ &        $75.29{\pm}0.27$ &  $70.02{\pm}0.45$ &            $58.53$ &                $0.201$ \\

\hline
Meta-graph & -- & -- &   $83.31{\pm}0.74$ &  \underline{\underline{$83.09{\pm}0.78$}} &          $5385.48$ &               $18.002$ & $75.27{\pm}0.46$ &   \underline{\underline{$68.34{\pm}0.76$}} &           $220.18$ &                $0.654$ \\
\hline

\multirow{ 2}{*}{HGFND} & METIS: & 500  &   \multicolumn{4}{c|}{out-of-memory}      &  $73.46{\pm}0.92$ &  $63.0{\pm}1.71$ &          $1429.13$ &                $1.886$ \\

& " & 1000   &   \multicolumn{4}{c|}{out-of-memory} &  $73.49{\pm}0.74$ &  $63.89{\pm}0.61$ &          $2401.68$ &                $3.248$ \\

\hline
Cluster-GCN & METIS: & $\ddagger$ &  $71.62{\pm}1.28$ &  $63.36{\pm}2.39$ &          $5851.02$ &               $15.212$ &  $73.15{\pm}0.57$ &  $61.8{\pm}0.75$ &           $236.44$ &                $0.638$ \\

\hline
\hline

\multicolumn{3}{|c|}{}& \multicolumn{4}{c}{\textbf{MM-COVID}}& \multicolumn{4}{c|}{\textbf{Health Release}} \\
 \cline{1-11} 
 
\multirow{7}{*}{HyperGraphDis} &METIS:  &20   &  $84.48{\pm}2.21$ &   $84.52{\pm}2.3$ &            $13.91$ &                $0.062$ &  $58.02{\pm}2.94$      &  $56.85{\pm}2.62$ &            $14.24$ &                $0.068$ \\
& " & 50   &  $86.63{\pm}2.43$ &  $86.79{\pm}2.43$ &            $16.54$ &                $0.066$ &  $59.04{\pm}3.09$      &  $58.91{\pm}3.13$ &            $14.27$ &                 $0.07$ \\
& "& 100  &  $86.46{\pm}1.35$ &  $86.45{\pm}1.46$ &            $18.61$ &                $0.075$ &  $55.69{\pm}1.46$      &  $55.65{\pm}1.47$ &            $15.15$ &                $0.068$ \\
& "& 500  &  $88.01{\pm}0.95$ &  $88.05{\pm}0.96$ &            $22.92$ &                $0.089$ &  $61.78{\pm}1.82$ & \bm{$61.7{\pm}1.93$} &    $17.35$ &                $0.079$ \\
& "& 1000 &   $89.5{\pm}0.74$ &  \bm{$89.48{\pm}0.66$} &  $24.99$ &                $0.094$ &  $61.68{\pm}2.12$      &  $61.56{\pm}2.25$ &            $18.69$ &                $0.084$ \\

% \hdashline
 & Infomap: & * &  $83.31{\pm}1.44$ &  $83.17{\pm}1.39$ &             $9.65$ &                $0.053$ &  $55.79{\pm}2.16$ &  $55.71{\pm}2.04$ &              $8.8$ &                $0.054$ \\

 & Louvain: & *  &  $83.81{\pm}1.5$ &  $83.79{\pm}1.57$ &            $10.17$ &                $0.056$ &  $59.14{\pm}0.98$ &  $59.07{\pm}0.96$ &             $8.98$ &                $0.047$ \\

\hline
Meta-graph & -- &  -- &  $85.47{\pm}1.98$ &   \underline{\underline{$85.41{\pm}2.02$}} &           $190.24$ &                $0.643$ & $54.42{\pm}2.07$ &  $53.66{\pm}2.11$ &             $8.59$ &                $0.054$ \\
\hline

\multirow{ 2}{*}{HGFND} 
&METIS: &500 &  $81.38{\pm}2.86$ &  $79.79{\pm}3.43$ &            $55.72$ &                $0.075$ &  $57.06{\pm}1.13$ &  $51.36{\pm}4.55$ &            $42.02$ &                $0.078$ \\

& " &1000 &  $84.03{\pm}3.12$ &  $83.06{\pm}3.7$ &            $81.52$ &                $0.116$ &  $56.45{\pm}1.57$ &   \underline{\underline{$53.95{\pm}3.53$}} &            $60.01$ &                $0.102$ \\

\hline
Cluster-GCN & METIS: & $\ddagger$ &  $68.84{\pm}1.59$ &  $56.15{\pm}2.06$ &           $175.94$ &                $0.458$ &  $50.0{\pm}3.75$ &  $41.69{\pm}3.16$ &            $24.39$ &                $0.073$ \\

\hline
\end{tabular}
}
\end{small}
\caption{HyperGraphDis versus other state-of-the-art methods. 
The best performance (in terms of F1 score) is in bold and is achieved by the HyperGraphDis across the four datasets. The second-best performance (by one of the other three methods) is underlined for each dataset. *Infomap \& Louvain: see Table~\ref{tab:social-meta-graph-statistics} for the number of communities produced by the community detection analysis. \textsuperscript{$\ddagger$}Cluster-GCN: the internal METIS partitioning is 10 clusters (mini-batch of 5 clusters) for the US Election and FakeHealth datasets, and 6 clusters for MM-COVID (mini-batch of 3 clusters).}
\label{tab:results_metis}
\end{center}
\end{table*}

\subsection{Main Results}

In this section, we delve into the results obtained from implementing our HyperGraphDis method across various datasets and compare them against existing state-of-the-art models. Specifically:

(1) Evaluation of HyperGraphDis on hypergraphs constructed by METIS versus on hypergraphs produced by community detection algorithms -- Infomap and Louvain~\cite{infomap, louvain}. Both approaches are applied to users' social networks as we presented in Figure~\ref{fig:hypergraph_pipeline}. Table~\ref{tab:social-meta-graph-statistics} presents the general statistics of the social networks. We experimented with varying cluster sizes (see Table~\ref{tab:results_metis}), ranging from 20 to 5000 (US Election and Health Story), all of which were derived using the METIS for graph partitioning. For MM-COVID and Health Release, the largest partition we conducted is 1000 due to the small size of their social networks (in terms of the number of users). Generally, the community detection methods generate clusters of users that are imbalanced in size, whereas METIS clusters are generally balanced. 

(2) HyperGraphDis versus state-of-the-art: (i) The vanilla version of Meta-graph; (ii) The HGFND, when applied on the hypergraphs generated by our methodology -- i.e., on the same hypergraphs that the HyperGraphDis has been tested; (iii) The Cluster-GCN, when applied on the meta-graphs -- i.e., the graphs of cascades that are produced by the Meta-graph approach. METIS partitioning plays an integral role in the functionality of the Cluster-GCN. So, in this experiment, METIS is applied directly to the graphs of cascades (i.e., the meta-graphs). This experiment evaluates the hypergraph construction that we propose.

\begin{figure*}[ht]
    \centering
\subfloat[][]{\includegraphics[width=0.38\textwidth]{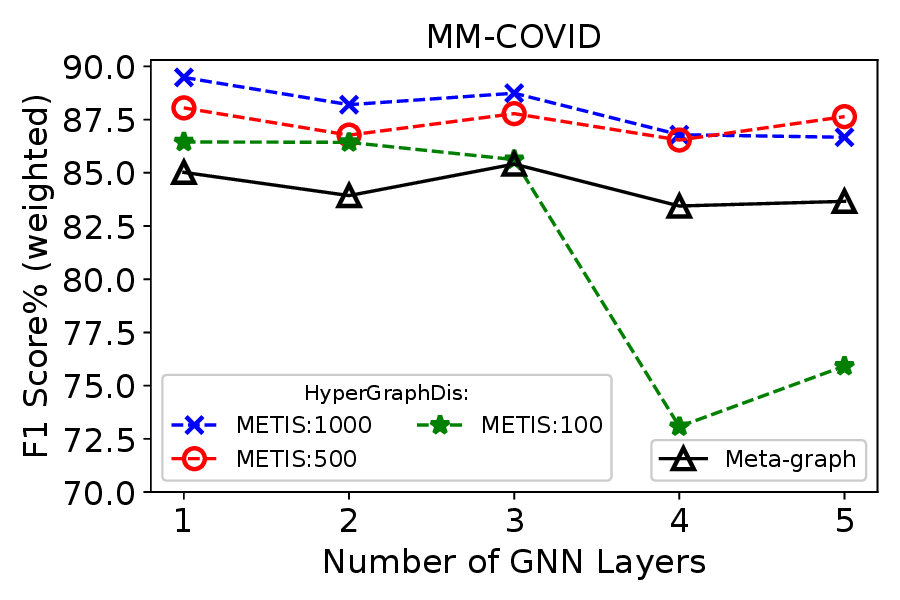}}
\hfil
\subfloat[][]{\includegraphics[width=0.38\textwidth]{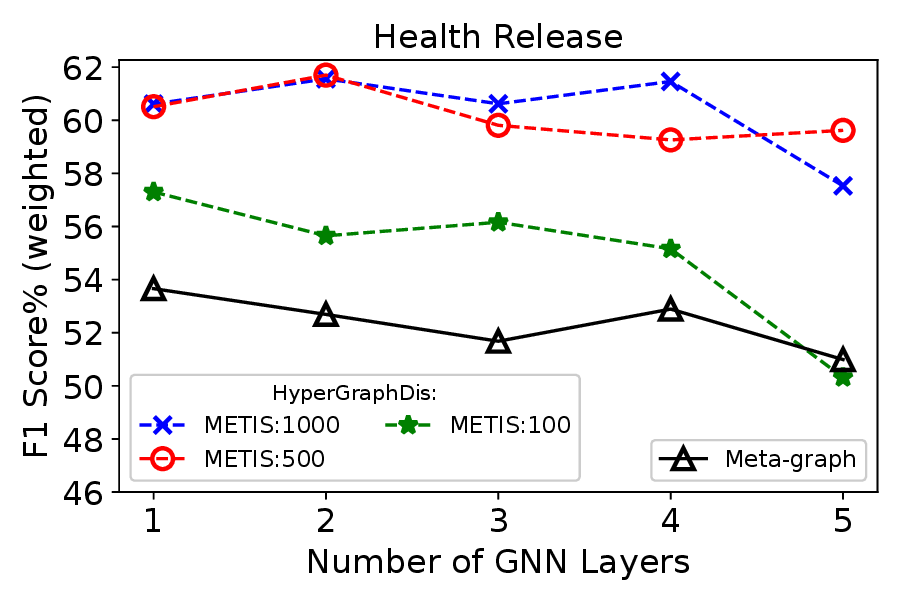}}
\hfil
\subfloat[][]{\includegraphics[width=0.38\textwidth]{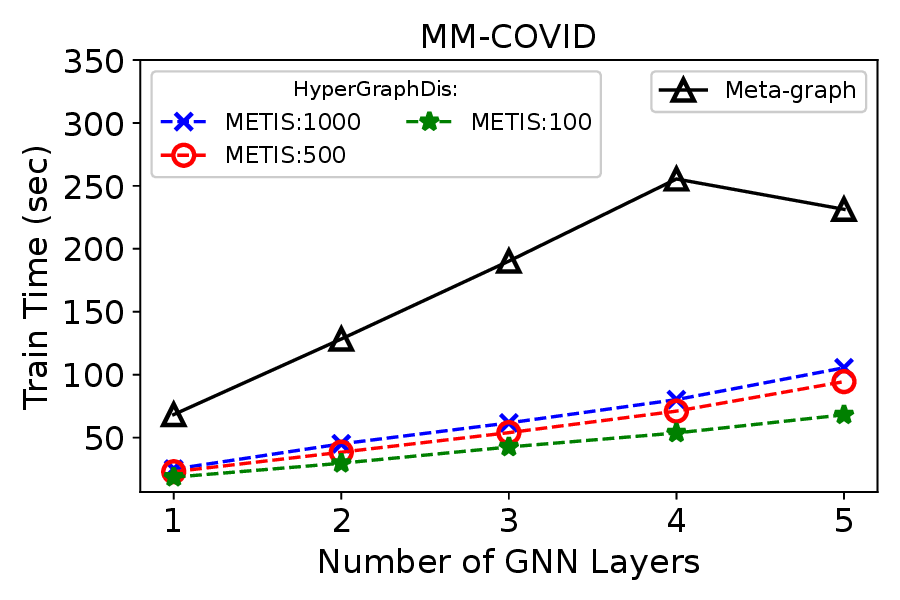}}
\hfil
\subfloat[][]{\includegraphics[width=0.38\textwidth]{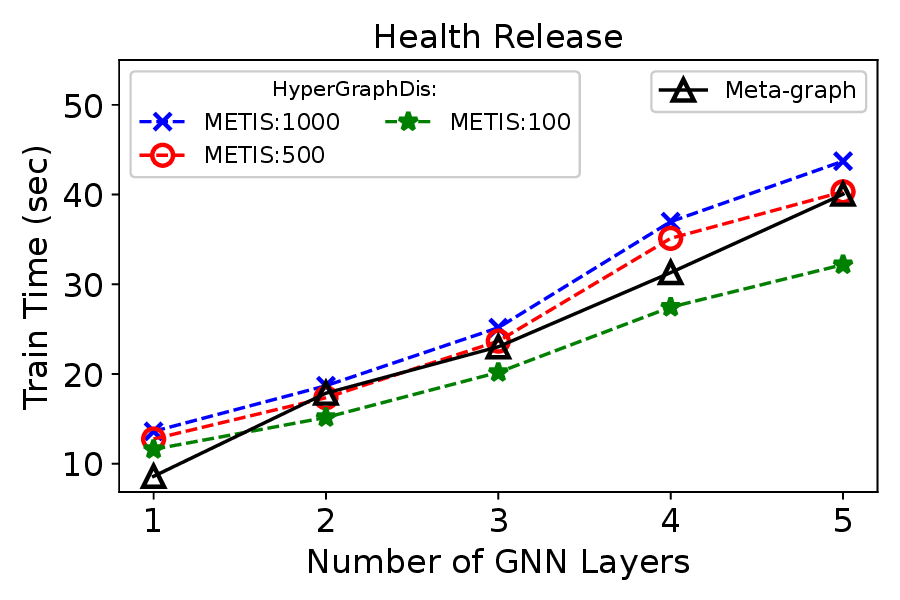}}
    \caption{Ablation analysis for HyperGraphDis and Meta-graph on MM-COVID and Health Release datasets. (a) \& (b): The GNN layers refer to the HypergraphConv layers in HyperGraphDis and to the GCNConv layers in Meta-graph. (c) \& (d): Train process time for the models evaluated in (a) and (b).  
    Five trials across all experiments.}
    \label{fig:ablation_1}%
    \end{figure*}

Regarding the number of GNN layers: 
\begin{itemize}
    \item HyperGraphDis: one HypergraphConv layer for US Election, MM-COVID, and Health Story; two layers for the Health Release. At the end of this Section, we present an ``Ablation Analysis'', where we show that more than one HypergraphConv layer does not increase the model's utility, in general (see Figure~\ref{fig:ablation_1}). 
    \item Meta-graph: one GCNConv layer for the US Election, three layers for the MM-COVID, two layers for the Health Story, and one for the Health Release. 
    \item HGNFD: two layers in the Hypergraph Attention network across all datasets.
    \item  Cluster-GCN: two GCNConv layers across all datasets. The internal METIS partitioning is 10 clusters (mini-batch of 5 clusters) for the US Election and FakeHealth datasets, and 6 clusters for MM-COVID (mini-batch of 3 clusters). 
\end{itemize}

%%%%%%%%%%%%%%%%%%%%%%%%%%%%%%%%%%%%%%%%%%%%
    
%%%%%%%%%%%%%%
    \begin{figure*}[ht]
    \centering
\subfloat[][]{\includegraphics[width=0.38\textwidth]{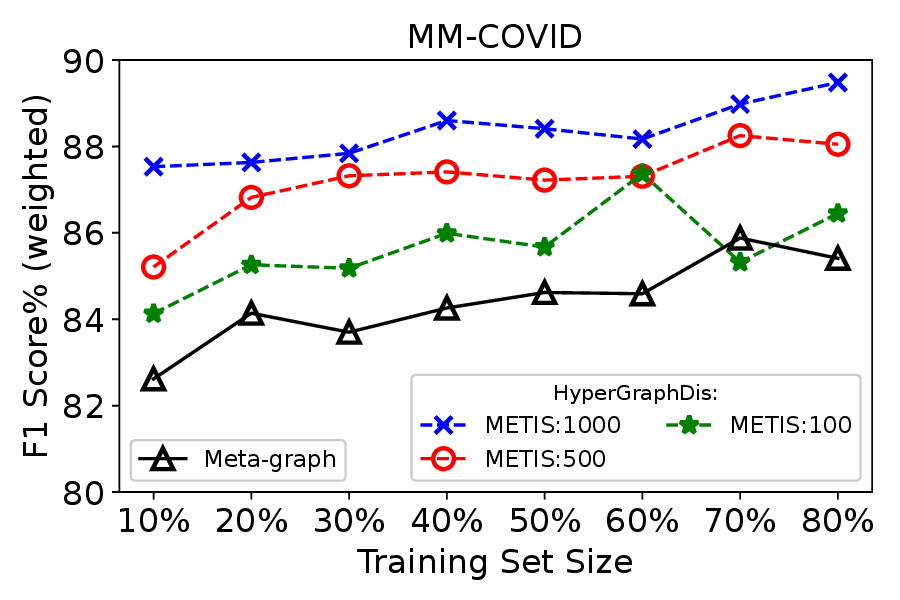}}
\hfil
\subfloat[][]{\includegraphics[width=0.38\textwidth]{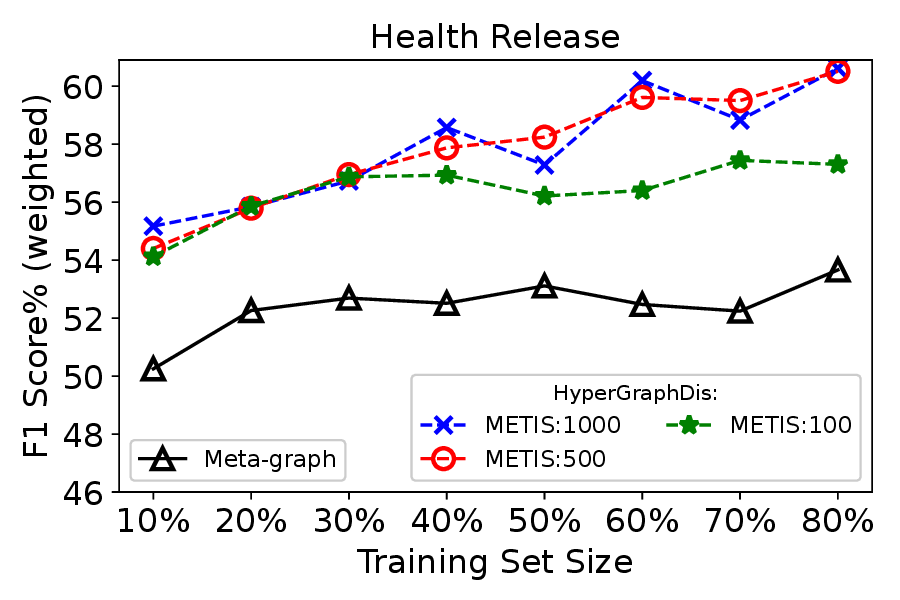}}

    \caption{Further evaluation of HyperGraphDis versus Meta-graph using the optimal number of GNN layers based on the results in Figure~\ref{fig:ablation_1}(a)\&(b). Namely, we use one HypergraphConv layer in MM-COVID and two layers in Heath Release for the HyperGraphDis. For the Meta-graph, we use three and one GCNConv layer in MM-COVID and Health Release respectively.}
    \label{fig:ablation_2}%
    \end{figure*}
% %%%%%%%%%%%%%%%%%%%%%%%%%%%%%%%%%%%%%%%%

The experimental results are summarized in Table~\ref{tab:results_metis}. 
The direct comparison against the previously described Meta-graph, HGFND and Cluster-GCN approaches depicts that HyperGraphDis displayed superior performance in weighted F1 score (average values on five trials). In MM-COVID the method achieves an impressive weighted test F1 score of around 89.5\%, outperforming the Meta-graph by approximately 4\%, HGFND by 6.4\%, and Cluster-GCN by 33\%. Turning to the largest dataset, US Election, HyperGraphDis yielded one of the most effective results with an F1 score of 85.73\%, outperforming the Meta-graph by 2.6\% and Cluster-GCN by 22.3\%. In the Health Story dataset, HyperGraphDis achieved 72.23\% F1 score outperforming Meta-graph, HGFND, and Cluster-GCN by 3.9\%, 8.3\%, and 10.4\%, respectively.

The overall results suggest that HyperGraphDis has broad applicability across different domains and information types, including political events and public health crises. These performance metrics highlight not only the method's robustness in analyzing politically charged disinformation but also its aptitude for handling imbalanced data, as reflected by the high F1 score. Finally, we found that the optimum performance on these three datasets was achieved with either METIS:500 or METIS:1000 clusters. The consistency in the optimal number of clusters may also indicate an underlying structure that is particularly amenable to hypergraph-based analyses. 

For the Health Release dataset, the results were notably different. The optimal performance was observed with a minimal cluster size of 500, delivering an F1 score of 61.7\%. It is important to note that this dataset is inherently less dense regarding its graph representation than the other datasets under consideration. The unique nature of the Twitter cascades in the Health Release dataset, being star graphs with a relatively small number of users (18.5K), likely contributed to this divergence. Despite the lower density and smaller user participation, HyperGraphDis still outperformed the other three methods, underscoring its flexibility and adaptability to various types of data structures.

%%%%
Beyond these primary metrics, we also observed a considerable reduction in both training and inference times when employing the hypergraph approach. This is a non-trivial benefit, especially considering the computational resources that large-scale data analysis demands. Specifically, significant improvements are noted in the processing time for both model training and inference processes. The training completion time is notably accelerated, ranging from 2.3 to 7.6 times faster than the second-best method (in terms of F1 score) across the four datasets (Meta-graph in US Election, Health Story, and MM-COVID; HGFND in Health Release). Similarly, the computational time showcases enhanced efficiency during inference, with an increase of 1.3 to 6.8 times compared to the second-best method. In summary, the results confirm that HyperGraphDis offers a compelling improvement over existing methods for identifying disinformation on Twitter. Not only does it deliver more accurate and nuanced results, but it also does so more efficiently, which is an essential quality for real-time or large-scale applications.

%%%
Finally, the classification performance of HyperGraphDis on the hypergraphs that have been constructed by the Louvain or Infomap graph partitions was quite low in both accuracy and F1 score. For instance, in the US Election dataset using the Infomap partition (2074 communities), we have achieved an accuracy of 78.37\% and an F1 score of 77.17\%. METIS methodology involves recursively partitioning a graph into more manageable subgraphs to identify densely connected regions. The process minimizes inter-cluster connections and balances load distribution, ensuring roughly equal cluster sizes. On the other hand, although Louvain and Infomap are highly efficient in identifying communities within complex networks, the communities they produce are not well interconnected and are not balanced in size. An indicator for that is that the training time on the hypergraphs produced by the Infomap is very low for the US Election Dataset, although the number of 2047 communities is quite high (see Table~\ref{tab:social-meta-graph-statistics}). HyperGraphDis (Infomap) achieved a 77.17\% F1 score in 113.77 seconds, whereas HyperGraphDis (METIS:500) achieved 85.73\%, but the training time took 1704.78 sec.

\subsubsection{Ablation Analysis}

We conducted an ablation analysis for HyperGraphDis and Meta-graph on MM-COVID and Health Release datasets to quantify (i) the effect of the number of HypergraphConv layers in HyperGraphDis and the number of GCNConv layers in Meta-graph; (ii) the effect of the training set size on both methods. We selected the two datasets where HyperGraphDis exhibited the best and worst performance (see Table~\ref{tab:results_metis}), and we also performed a comparative analysis against the Meta-graph (the second-best method). The ablation analysis for HyperGraphDis was performed on the hypergraphs constructed by METIS for different numbers of clusters (ranging from 100 to 1000).

Figure~\ref{fig:ablation_1} presents the average values (from 5 trials) of the weighted test F1 score. In MM-COVID, the experiments with different numbers of HypergraphConv layers (from 1 to 5) indicate that applying several layers does not increase the utility of the model (Figures~\ref{fig:ablation_1}(a) \& (b)). In Health Release, a model having two HypergraphConv layers achieved a slightly higher F1 score. Regarding the Meta-graph, a 3-layer model for MM-COVID and a 1-layer for Health Release depicts the highest performance in terms of F1 score. From Figures~\ref{fig:ablation_1}(c)\&(d), we observe that in MM-COVID, the training process time of the Meta-graph is heavily affected by a large number of GCNConv layers. This illustrates the limitations of the Meta-graph approach. The meta-graph structure for MM-COVID is very dense (see Table~\ref{tab:social-meta-graph-statistics}); consequently, the message-passing process of a model with several GCNConv layers will not be may not be efficient in terms of training process time. 

We also experimented with varying training set sizes, ranging from a train-test split of 10/90 to 80/20. The difference in performance between the two methods remains approximately the same across all training sizes, indicating the superiority of HyperGraphDis (METIS:500 \& 1000). Figure~\ref{fig:ablation_2}(a) shows that the HyperGraphDis (METIS:1000) in MM-COVID, a training set size of around 40\% is adequate to produce a model that can achieve an F1 score close to 89\%. In Health Release (Figure~\ref{fig:ablation_2}(d)), a training set size of 60\% is close to optimal.   

\subsubsection{Further Discussion on HyperGraphDis Performance} 

In this study, we evaluated the HyperGraphDis method using the MM-COVID and Health Story datasets. Other studies that used these two datasets are the following:

\textbf{MM-COVID} In \cite{sota-fake-news-detection}, a meta-analysis explored state-of-the-art Machine Learning techniques for \textit{text-based} fake news detection. Models, including Random Forest, CNN-based, LSTM-based, and BERT-based, underwent formal evaluation on benchmark datasets, such as MM-COVID. Notably, the BERT-based model achieved the highest F1 score (79.4\%), while Random Forest had the lowest score (63.3\%). The HyperGraphDis, with an F1 score of 89.48\%, showcases a significant 10-26\% improvement over conventional text-based methodologies. This performance advantage underscores the value of incorporating social network graph data and user-related features alongside text-based features in enhancing fake news detection tasks. 

\textbf{Health Story} \textit{SOMPS-Net framework}~\cite{somps-net}, comprising two essential components: the Social Interaction Graph and the Publisher and News Statistics. Tested with the Health Story dataset, it achieved a 72.7\% accuracy and 79.6\% F1 score in the realm of fake news prediction. We tested HyperGraphDis using the same dataset, although we observed slightly diminished results regarding the F1 score, specifically an accuracy of 75.21\% and an F1 score of 72.23\%. We attribute this decline in performance to the inherent characteristics of the cascades within the Health Story dataset, which tend to be either extremely small or predominantly structured like star-like graphs. This characteristic leads to embedding vectors with significant similarity, impacting the overall effectiveness of our method.

\subsubsection{Limitations} 
Although HyperGraphDis is computationally efficient (in terms of training and inference process time), making it suitable for disinformation detection in large datasets, there are still some limitations. First, a large number of tweets is needed to construct the social network -- a crucial component of the HyperGraphDis pipeline (see Figure~\ref{fig:hypergraph_pipeline}). The social network -- the user-user interaction graph -- is constructed by the interaction between the users (replies, retweets, mentions). For the produced social network to be valuable, it needs to have a rich graph structure, which heavily depends on the number of tweets collected. Secondly, in the current version of HyperGraphDis, we compute the users' graph representation (user embeddings) on each enhanced subgraph (per cascade) by applying the DeepWalk algorithm (see Figure~\ref{fig:cascades_vector_construction}). This approach increases the overall complexity of the pipeline construction. In a future study, we aim to address this problem by incorporating "global embeddings" -- i.e., the users' graph representation will be computed once for the entire social network or in a small region around each user. Community detection or graph sampling algorithms may assist us in this direction.

%%%%%%%%%%%%%

\begin{table}
\begin{small}
\begin{center}
\adjustbox{max width=\columnwidth}{%
\begin{tabular}{lllll}
    \hline
     & ML Model  & Acc & F1 Score & Features\\
    \hline
\multicolumn{5}{l}{ \textbf{MM-COVID} } \\ \hline
 \multirow{4}{2cm}{\makecell{Text-based ML methods \\ \cite{sota-fake-news-detection}}} 
& Random Forest & 87.6\% & 63.3\%   & \multirow{4}{2cm}{news content} \\
& CNN & 88.1\% & 67.3\%  \\
& LSTM & 90.4\% & 75.0\% \\   
& BERT & 92.5\%  & 79.4\% \\ 
\hline
    \multicolumn{5}{l}{\textbf{Health Story}} \\ \hline
     {\makecell{SOMPS-Net \\ \cite{somps-net}}}  &GCN  & 72.7\% & 79.6\% & \makecell{User, network, \\news content}\\ 
\hline  
\end{tabular}
}
\caption{Other disinformation detection methods on MM-COVID and Health Story datasets.}	
	\label{tab:other_solutions}
\end{center}
\end{small}
\end{table}
%%%%%%%%%%%%%%%%%%%%%%

\section{Conclusion}
\label{conclusion}
The problem of disinformation on social media is a pressing issue with substantial social, economic, and political implications. In this paper, we introduce a novel method, HyperGraphDis, designed to combat the spread of disinformation by analyzing Twitter data through the lens of a hypergraph data structure. Hypergraphs offer several advantages over traditional graph-based methods. They provide enhanced scalability, which is crucial for efficiently analyzing vast social media datasets. We applied HyperGraphDis to diverse datasets, including health and socio-political domains to validate our approach. The method's versatility was evident, showcasing its potential across various applications. Comparative tests against state-of-the-art approaches revealed significant accuracy and F1 score improvements, crucial for imbalanced datasets common in disinformation detection. Furthermore, our method displayed improvements in computational cost and run times, reinforcing its advantages in terms of scalability and efficiency.

\section{Ethics}
This work followed the principles and guidelines on executing ethical information research and using shared data~\cite{ethics}. The suggested methodology complies with the GDPR and ePrivacy regulations. We use Twitter datasets that have already been published. We did not use or present any identifiable user information from the datasets (e.g., Twitter user IDs). We applied text preprocessing to clean the tweets from any information that could identify specific Twitter accounts. 
Finally, we implemented and executed the experiments locally on our machines, so we did not upload any of the datasets to the cloud.

\section{Acknowledgments}
This work has been funded by the European Union under the projects MedDMO (Grant Agreement no. 101083756) and INCOGNITO (Grant Agreement no. 824015).

\bibliography{HyperGraphDis}

\end{document}